\documentclass[9pt,conference]{IEEEtran}
\usepackage[preprint]{waspaa25}
\usepackage{bm} 

\title{Toward a Sparse and Interpretable Audio Codec}

\name{John Vinyard}
\address{
  Austin TX, USA \\
  \href{mailto:john.vinyard@gmail.com}{john.vinyard@gmail.com}
}

\begin{document}

\maketitle

\begin{abstract}
  Most widely-used modern audio codecs, such as Ogg Vorbis and MP3, as well 
  as more recent ``neural'' codecs like Meta's Encodec\cite{defossez2022highfidelityneuralaudio} 
  or the Descript Audio Codec\cite{kumar2023highfidelityaudiocompressionimproved}
  are based on block-coding; audio is divided into overlapping, 
  fixed-size ``frames'' which are then compressed. While they often yield excellent 
  reproductions and can be used for downstream 
  tasks such as text-to-audio, they do not produce an intuitive, 
  directly-interpretable representation.  In this work, we introduce a 
  proof-of-concept audio encoder that represents audio as a sparse 
  set of events and their times-of-occurrence. Rudimentary physics-based 
  assumptions are used to model attack and the physical resonance of 
  both the instrument being played and the room in which a performance 
  occurs, hopefully encouraging a sparse, parsimonious, and 
  easy-to-interpret representation.
\end{abstract}

\section{Introduction}
\label{sec:intro}

This work imagines a future audio codec where some types of musical 
composition could take place directly in the audio codec space. 
The text-to-audio paradigm works well for non-musicians creating background 
music for movies, advertisements or social media content, but it is 
our view that experienced musicians and composers work in a 
``space'' not fully captured by language alone.  We theorize that they will 
continue to prefer a finer-grained representation that provides 
interpretability and control at multiple scales.

This work does not seek to produce a generative model of musical audio, 
but could serve as the underlying encoding on which generative 
models are trained. It is the authors' intuition that models 
trained on this quasi-symbolic representation might have a 
much more thorough ``understanding'' of the content they produce, 
given the point-cloud-like nature of the signal. Instead of 
predicting the next arbitrarily-sized frame, the generative model 
would be predicting the relationships between ``events''. Long-term 
coherence in musical generation continues to be a challenge, and we 
speculate that many models spend large shares of their capacity 
learning to reproduce physical resonance, reducing the share 
that may model the human forces that drive them in interesting, 
musical directions.

All model and experiment code is implemented using the PyTorch\cite{NEURIPS2019_9015} Python library and 
can be found on GitHub\footnote{\href{https://github.com/JohnVinyard/matching-pursuit}{https://github.com/JohnVinyard/matching-pursuit}}. 
Audio examples can be heard at this https URL\footnote{\href{https://blog.cochlea.xyz/sparse-interpretable-audio-codec-paper.html}{https://blog.cochlea.xyz/sparse-interpretable-audio-codec-paper.html}}

\begin{figure}[t]
  \centering
  \centerline{\includegraphics[clip, trim=0cm 8cm 0cm 8cm, width=1.00, width=\columnwidth]{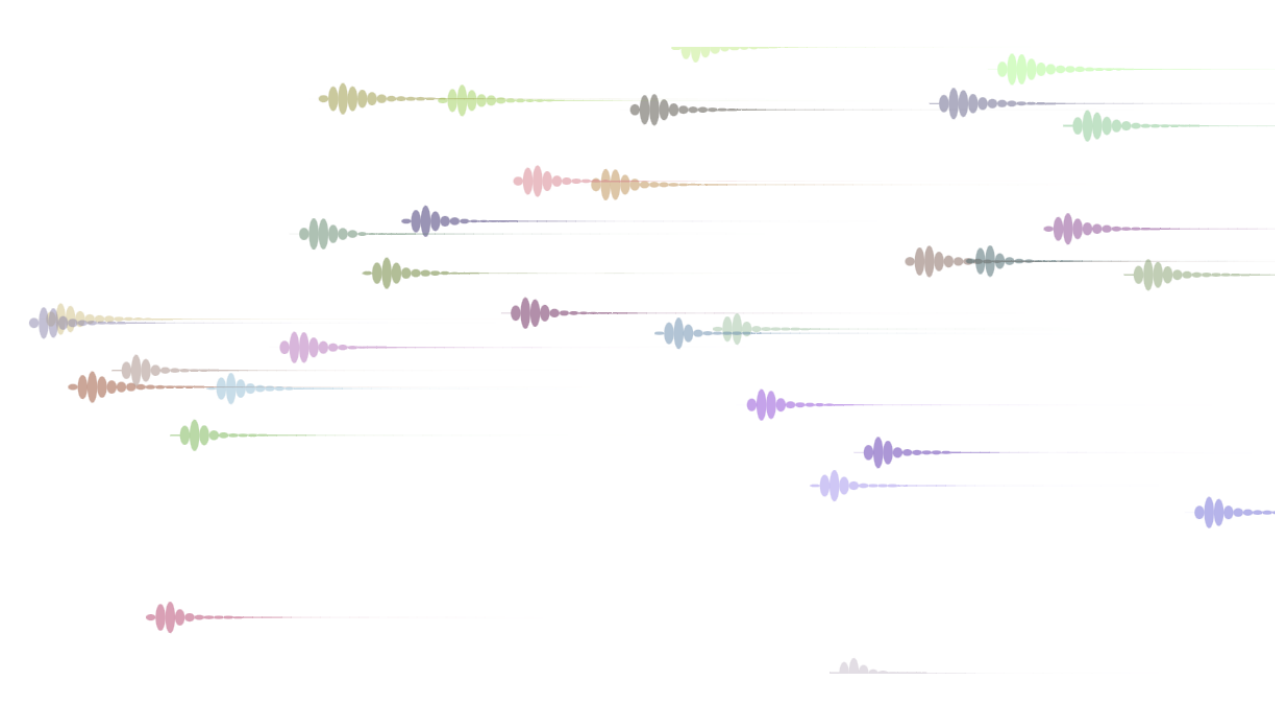}}
  \caption{
    In this visualization of the codec representation, we see that events can overlap and vary in length.  
    Time is along the x-axis, event positions are along the y-axis and event colors are chosen by 
    applying t-SNE\cite{vanDerMaaten2008} to the set of 32 event vectors, targeting a single dimension for the y-axis and 
    three dimensions to represent an RGB color value.
  }
  \label{fig:audio_events}
\end{figure}

\section{Prior Work}
\label{sec:prior_work}

We take inspiration from a wealth of previous works dealing with sparse and interpretable
representations of audio signals.  Matching pursuit\cite{10.5555/890205} is an iterative 
algorithm that decomposes a signal into a sparse set of representative ``atoms''.  It produces
a representation of audio signals similar to the one used in 
granular synthesis\cite{xenakis_formalized_1971}, which represents audio as point-cloud-like
structure of audio ``grains'' arranged in time.  While the notion of ``events in time'' feels
intutive to many of us, both approaches traditionally require hundreds or thousands 
of ``atoms'' or ``grains'' to accurately reconstruct a signal, These audio quanta are often orders of shorter than 
the time scale on which we typically conceive of audio events.  Furthermore, matching pursuit, at least when 
operating naively in the time domain can spend capacity on perceptually irrelevant details and 
struggle to reproduce noisy signals in a convincing way.  In recent years, attempts have 
been made to use longer atoms and paramterize dictionaries of atoms in more meaningful ways\cite{reds2017}.
In this work we hope to continue the trend toward fewer and more meaningful ``atoms'' or ``events'' 
that exist at scales native to a human musician.

Unsupervised audio source separation, such as 
in\cite{schulzeforster2023unsupervisedmusicsourceseparation} seeks to separate mixtures of
instruments or voices into distinct tracks, or ``stems'' as they are often referred to in 
music production, while unsupervised music transcription\cite{NIPS2014_e51e118b} seeks to infer
a musical score, or set of instructions sufficient for a musician to reproduce the piece.  
Other recent works seek to produce meaningful latent representations by encouraging sparsity 
as part of the training objective\cite{lisboa2024spikingmusicaudiocompression}
In this work, we seek to infer a score of sorts, where each encoded note or event contains 
enough information to create a perceptually indistinguishable reproduction of the event.

Finally, source-excitation synthesis\cite{fant1960acoustic} models natural sounds as injections 
of energy into a system, often represented by a burst or pulse-train of white noise, and the
resonance of that system in response to the event, often represented by one or more time-varying
filters that emulate the resonant modes of the system along with any deformations of the system that occur as it 
resonates.  In this work, we use this synthesis technique as the basis of the event decoder.  Empirically, this 
strong inductive bias seems to aid in the promotion of sparsity (few events) and to provide us with
intermediate states that are open to inspection and interpretation.

\section{Audio Codec Details}
\label{sec:model}

Our proposed audio encoding consists of a sequence of two-tuples, each consisting of
a scalar event time and the parameters required for the decoder to render
the event, transforming it into ``raw'' PCM audio samples.  In addition to
straightforward seeking and slicing, this encoding also makes it possible to 
filter a subset of events in a particular audio slice according to some 
criteria, an operation that is less straightforward in a block-coding scheme.

\subsection{Compression Rate}
\label{ssec:compression_rate}

While our main aim in this work is a representation 
that is sparse, interpretable and easy-to-manipulate, a
compressive representation is one important aspect of a successful codec and should be discussed.
For our experiment, the encoder is run for a fixed number of steps, 32 in our case,
and produces a two-tuple of time-of-occurrence and 32-dimensional event vector.  
Each event time can be stored as a single scalar value.
Because we encode ${2^{16}}$ samples, or around 2.98 seconds of audio at a time, 
at a 22050hz sampling rate, we arrive at:

\begin{equation}
  \label{eq:wave_equation}
    \frac{2^{16} samples}{(32 * 32) + 32} = \backsim62x
\end{equation}

\section{Model}
\label{sec:model}

\subsection{Audio Representation}
\label{ssec:audio_representation}

In this work, we seek an audio representation that can be iteratively decomposed, 
without wasting capacity on perceptually-irrelevant details.  As a concrete example, 
removing energy from a time-domain representation of noise can be very difficult, as the
only way to do so is to match the signal \emph{exactly} even though small time and/or
frequency shifts would be imperceptibile to the common listener.

We choose the widely-used STFT magnitude spectrogram representation of audio, with a 
window size of 2048 and a hop size of 256.  The 75\% overlap means that we can discard
phase but still recover \emph{most} perceptually relevant details, including phase relationships.

While the model only encodes segments of ${2^{16}}$ samples at a time, 
it \emph{analyzes} segments of ${2^{17}}$, encoding and removing only events that 
begin in the first half of the audio segment.  This enables us to implemenent a streaming 
encoder.  The model's input at each step is an STFT magnitude spectrogram with 1025 real-valued coefficients 
and 512 ``frames'' in the time dimension.

\subsection{Encoder}
\label{ssec:encoder}

In this work, the encoder is paramaterized as an anti-causal, 
dilated convolutional network with a kernel size of 2 at 
each layer and dilation sizes of [1, 2, 4, 8, 16, 32, 64].

In our experiments, the encoder runs for a fixed number of steps (32).  At each step,
it transforms the residual spectrogram into a tensor with 32 channels and 512 frames, 
the same time dimension as the input spectrogram.  The ``residual'' spectrogram at the first step
is simply the STFT magnitude spectrogram computed from the input audio.

The encoder then selects a \emph{single} event vector along the time dimension, 
setting all other 32-dimensional vectors to zero.  
When we take the norm of each position along the time dimension, we derive a one-hot
vector that will be used for the coarse-grained timing of the event when it is 
``scheduled'' after the event decoder has generated the event.

Once generated and scheduled, the STFT magnitude spectrogram of the 
newly-generated event is computed and subtracted from the residual spectrogram, 
preparing the model to begin the next iteration.

While our initial experiments use a fixed number of encoding steps, we believe that we
will be able to devise more intelligent stopping conditions in future versions of this work 
that take advantage of the variable information density across different audio segments.  
One simple and obvious possibility is to choose a small norm as the threshold at which
the sound would no longer be audible to listeners.

\begin{figure}[t]
  \centering
  \centerline{\includegraphics[width=\columnwidth]{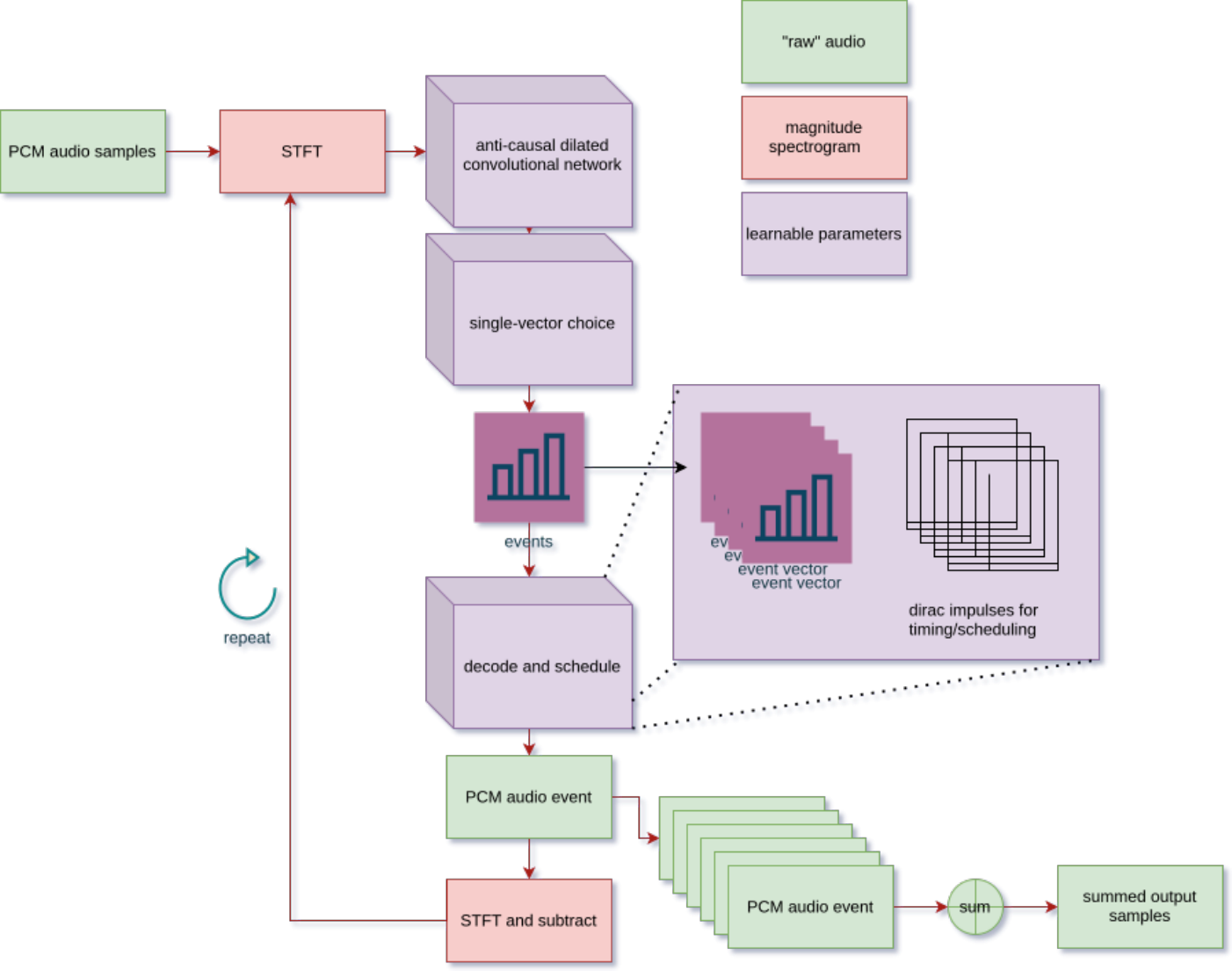}}
  \caption{High-level model architecture.  The encoder and decoder 
  work together to incrementally remove energy from the input 
  representation, an STFT-based magnitude spectrogram.  
  At each step, the encoder produces a single event vector and 
  time-of-occurrence.  The decoder generates an audio 
  event, positions it in time, and subtracts it from the input representation.}
  \label{fig:model_architecture}
\end{figure}

\subsection{Streaming Algorithm}
\label{ssec:streaming_algorithm}

To enable a streaming encoding algorithm, the encoder masks the second-half of its 
output just before choosing the next event vector to be decoded and removed.  
This means that the encoder is always choosing events that begin in the first-half of
its input, but the events it chooses may into the second half.

To enforce this focus on the first half of the analyzed signal, we mask encoder
output before choosing an event, \emph{and} multiply the second half of the signal 
by a linear gradient beginning at 1 and ending at 0 which extends from 
sample ${2^{16}}$ to sample ${2^{17}}$.  This means that model is penalized less
for failures to remove energy from the second half of the signal and that it does not
attempt to produce overly-long events that stretch far into the second-half.

\subsection{Decoder}
\label{ssec:decoder}

\begin{figure}[t]
  \centering
  \centerline{\includegraphics[clip, trim=0cm 8cm 0cm 2cm, width=1.00, width=\columnwidth]{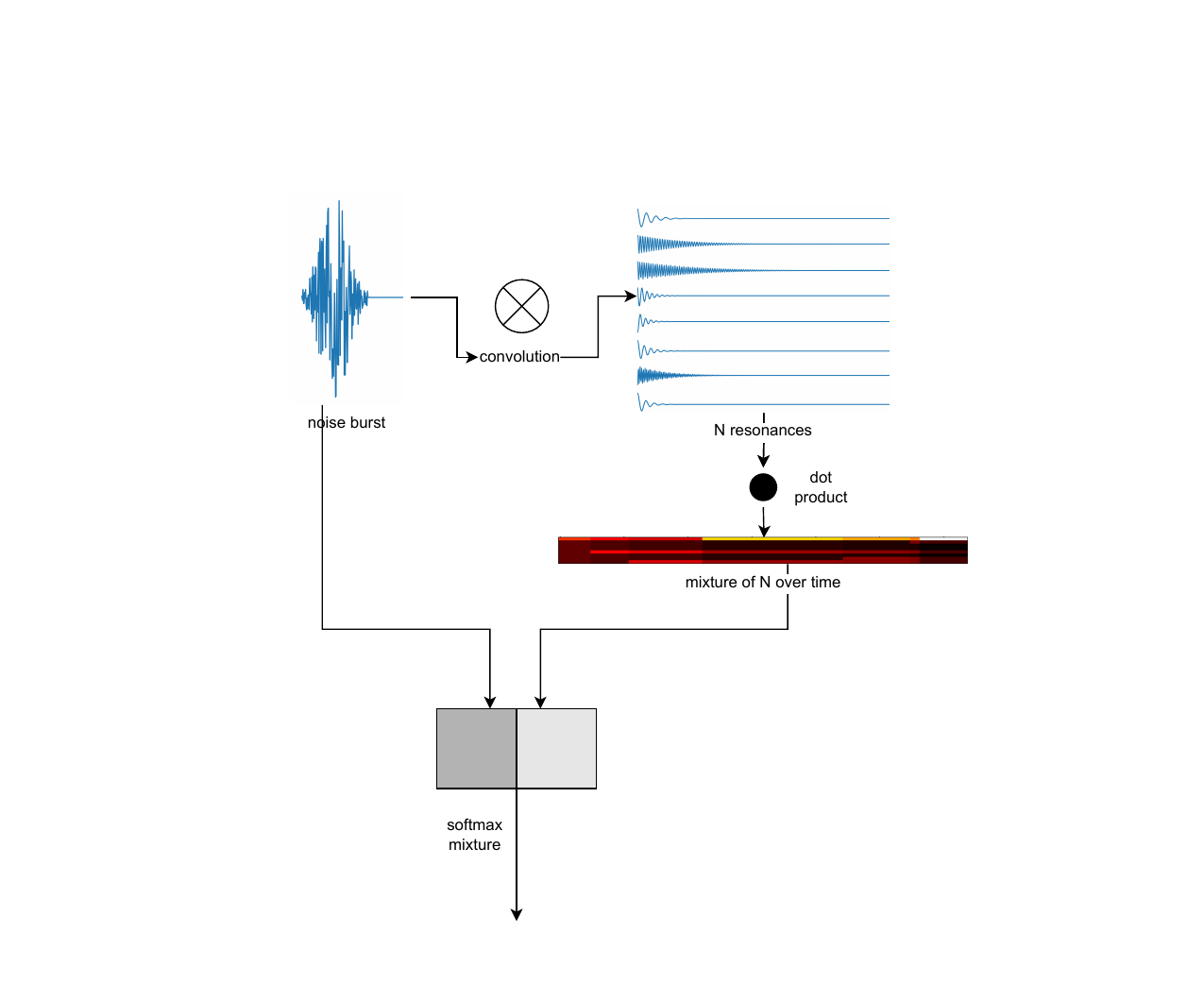}}
  \caption{Here, we can see the anatomy of a decoder block, which performs 
  something like source-excitation synthesis.  A burst of noise is convolved with a number of decaying resonances.  
  A time-varying mixture interpolates between the different resonances over time, after which the original impulse 
  and the resonant signal are mixed together before being output.}
  \label{fig:decoder_block}
\end{figure}

Instead of a more traditional convolutional upsampling network, we choose a 
source-excitation model for the event decoder in the hope that it encourages both
sparser and easier-to-interpret events.

In Figure \ref{fig:decoder_block}, we can see a diagram of a single decoder block.  
Each block has a few distinct parts with clear interpretations.  Inputs to each block
are generated by a small, ``multi-head'' MLP whose input is the original 32-dimensional 
event vector.

The input to each block is a source signal, which takes the form of one or more 
bursts of noise that represent the injection of energy into the system.  The noise burst
is convolved with a number of decaying resonances, represented in our work as relatively 
long FIR filters composed of exponentially decaying sinusoids.  Each convolved resonance is then multiplied 
by a time-varying mask and summed, emulating a filter with time-varying parameters, or in 
a more physical interpretation, the deformations applied over time to a resonating object.

We refer to the number of resonances chosen for a particular block as its expressivity.
Some simple physical resonances, such as a tuning fork, might only require that expressivity=1, 
while others, such as a vibrato violin note, might require a larger expressivity number.

Finally a two-channel gain is applied to the original source and the time-varying, 
filtered-signal, akin to a weighted skip connection.

These decoder blocks can be stacked, with the output of one block serving as the 
input or ``impulse'' fed to the next.

For this set of experiments, we choose three decoder blocks, with the last 
layer being fixed and given an expressivity value of 1.  This block is initialized 
with a set of freely-available room impulse responses\cite{voxengoir} that are commonly 
used to produce reverb effects.  This makes it \emph{possible} to disentangle instrument 
and room resonances.

For coarse-grained scheduling, the audio generated by the decoder is first 
convolved with the one-hot vector produced by the encoder.  Fine-grained, 
sample-level scheduling is achieved using a frequency-domain time-shift, 
represented by a single scalar value, that is generated from the event vector
by one of the decoder's MLP ``heads''.

\subsection{Model Size}
\label{ssec:model_size}

The model used in our experiments has a total of 45.1M parameters and occupies 
200MB when serialized to disk.  We feel that even this small model shows promise 
and that further experiments with larger models are warranted.

\section{Experiment}
\label{sec:experiments}

Our primary experimental goal is to show that we can encode a large set of 
``natural'' (non-synthesized) musical audio with diverse instruments and 
recording conditions into a representation that enjoys sparsity, interpretability 
and good reproduction quality.

\subsection{Data Set}
\label{ssec:data_set}

We train our model on the MusicNet dataset\cite{thickstun2017learning}, an open dataset containing ~33 hours 
of classical music, recorded in diverse spaces using a range of 
recording equipment of varying quality.  The recordings often include leading 
silence, trailing applause, and incidental human sounds throughout 
(coughs, movement, etc).

While the dataset as a whole includes fine-grained score information about each musical piece, 
we do not use this component of the dataset, using \emph{only} the audio signals to learn our 
unsupervised model.

\subsection{Training Algorithm}
\label{ssec:training}

During training, we repeatedly draw segments of audio that are approximately 5.94 seconds in 
length from the Music Net dataset at random, with 
length ${2^{17}}$ samples at 22050hz sampling rate and a batch size of two.  

We train using a single NVIDIA GeForce RTX 3060 for approximately 76 hours.

Each batch is passed through the encoding and decoding process for a fixed number of steps, 32, in 
our case, and then an iterative loss is computed, encouraging each step to have removed as much 
energy (expressed as the L1 norm) from the signal as possible.  This loss is minimized using 
the Adam optimizer with a learning rate of ${1e^{-4}}$.  
Empirically, we found that in early stages of training, the model 
tended to simply output silence as a pathological local minima.  
Empirically, the relatively snall learning rate seemed to help move beyond these early-stage 
pathologies, but a learning-rate schedule that increases once the model
has stabilized might yield better results overall.

\subsection{Loss Functions}
\label{ssec:loss_functions}

The loss function is also iterative and greedy.  Using the same 
magnitude spectrogram representation analyed by the encoder, 
the loss function attempts to \emph{maximize} the reduction in the 
L1 norm of the spectrogram at each step.  Prior to computing 
the loss, the input events are sorted in order of descending 
L1 norm, such that the event with the most energy is subtracted 
first and the event with the least energy is subtracted last.

\section{Interpretable and Manipulatable Representation}
\label{sec:interpretable_representation}

In this section, we discuss the multiple scales of 
interpretability and manipulatibliity provided by the propsed codec.  We intuit that 
these properties will be appealing to musicians and sound designers, and speculate that 
these features may prove just as important as compression rates under 
some circumstances.

\subsection{Events and Times-of-Occurrence}
\label{ssec:events_and_times}

As shown in Figure\ref{fig:audio_events}, relatively few events are required 
to reconstruct musical audio segments.  Information density and temporal structure
is apparent at-a-glance.  At this scale, it is possible to preview individual events, 
translate them in time, and add or remove events based on some criteria.

\subsection{Events Vectors}
\label{ssec:event_vectors}

\begin{figure}[t]
  \centering
  \centerline{\includegraphics[clip, trim=0cm 5cm 0cm 5cm, width=1.00, width=\columnwidth]{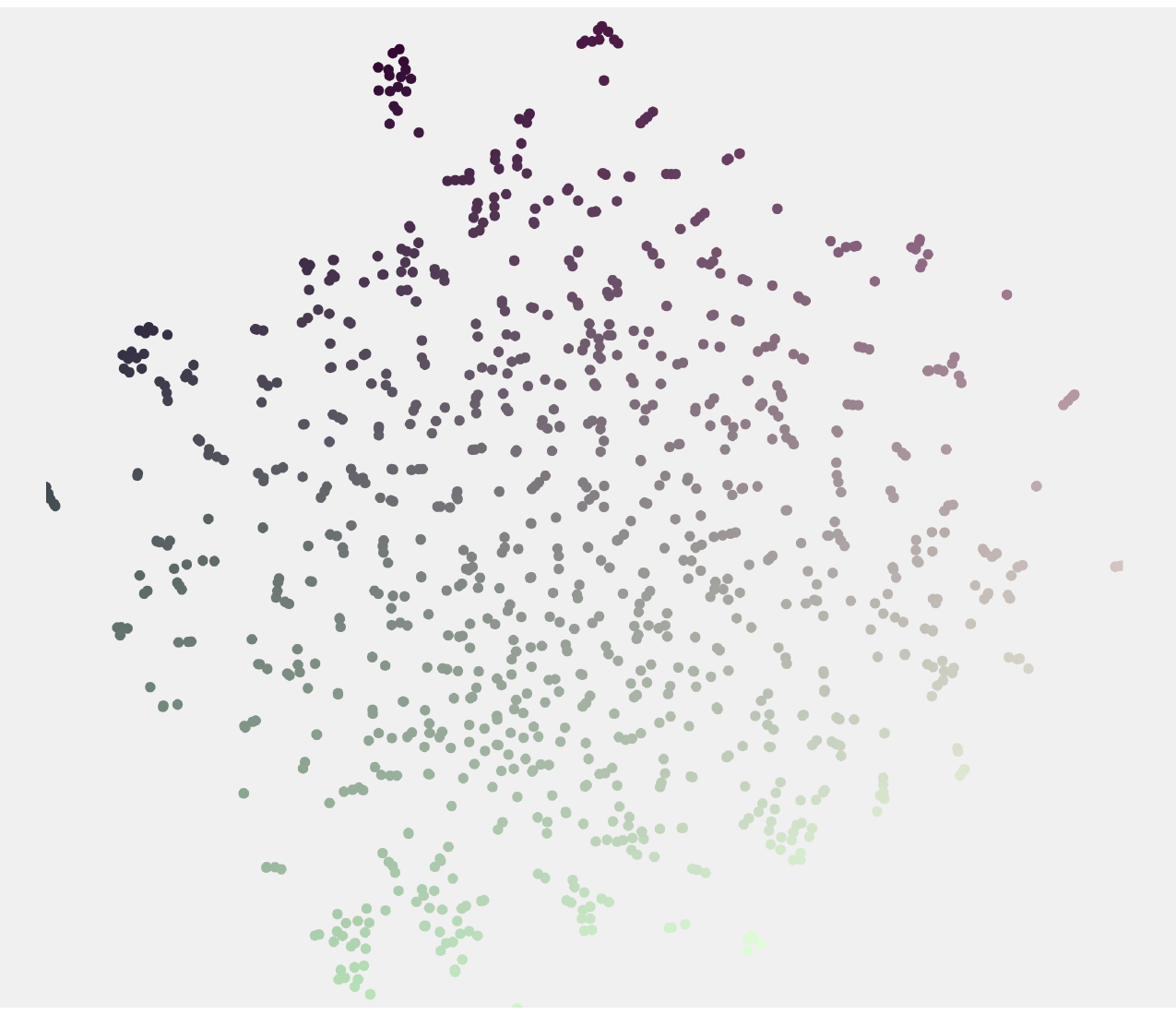}}
  \caption{
    This scatterplot shows event vectors from a large number of audio 
    segments mapped onto a 2D-plane using t-SNE\cite{vanDerMaaten2008}.  
    Exploring nearby neighbors can locate variations of a query event.  
    An interactive version of this scatterplot can be 
    explored \href{https://blog.cochlea.xyz/scatter.html}{here}.
  }
  \label{fig:event_scatterplot}
\end{figure}

Aside from a coarse-grained event time, the low-dimensional event vectors 
encode all information required to produce an audio event.  This work does
not explore the latent space learned by the encoder in-depth, but an interactive 
two-dimensional map of event vectors (see Figure \ref{fig:event_scatterplot}) 
demonstrates that a nearest-neighbors exploration can locate interesting variations
on a particular event.

\subsection{Decoder Interpretation}
\label{ssec:decoder_interpretation}

\begin{figure}[t]
  \centering
  \centerline{\includegraphics[clip, trim=0cm 20cm 0cm 5cm, width=1.00, width=\columnwidth]{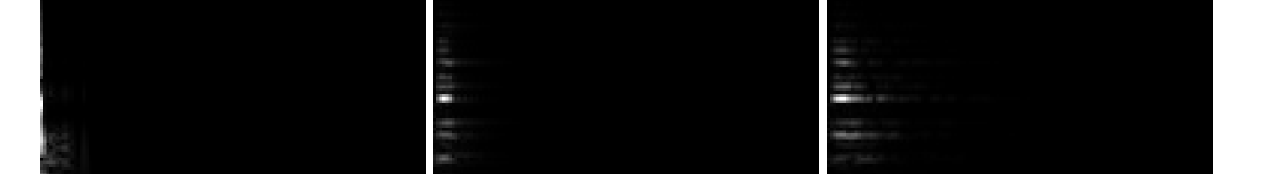}}
  \caption{
    Here we can see intermediate stages of the source-excitation-based decoder at work.  
    From left to right, we see A. a spectrogram of the initial noisy 
    impulse B. a spectogram of the noisy impulse convolved with the chosen decaying resonance and C. the result from 
    step B. convolved with a room impulse response.
  }
  \label{fig:event_detail}
\end{figure}

While the high-level model architecture is decoder-agnostic, our choice of
a source-excitation-inspired decoder in this first experiment means that further 
interpretation and manipulation is possible by inspecting intermediate decoder states.
In Figure \ref{fig:event_detail} we see spectrograms of the initial impulse, the resonance, 
and finally the selected room impulse-response for a single event.  While this
work does not include an in-depth exploration of decoder interpration and manipulation, it's possible
that different impulses, resonances, or room impulse responses could be chosen, 
while other aspects are held constant in order to subtly or profoundly influence
the rendering of a particular event.

\section{Conclusion}
\label{sec:conclusion}

In this work, we propose an audio codec optimized for 
interpretability and ease-of-manipulation and then discuss a simple, 
small-scale reference implementation of an encoder and decoder network.  
We find that while subjective reproduction quality falls short in 
this iteration, the results achieved with a relatively small network are
encouraging, and the properties of the codec that promote intuitive 
understanding of the audio content are worth further study.

\section{Future Work}
\label{sec:future_work}

Given the encouraging small-scale results, we feel that there are several
future experiments ripe for exploration.

\subsection{Perceptually-Inspired Losses}
\label{ssec:better_perceptual_losses}

Our sense is that perceptual audio losses are an under-explored area, and
that adversarial losses are often introduced to account for a fundamental mismatch 
between what commonly-used audio representations measure and what is 
perceptually meaningful to humans.  While the magnitude spectrogram 
representation used in this work may largely solve the problem of 
dealing with perceptual invariances around band-limited noise, it does not address higher-level and 
more complex invariances in human auditory perception, as discussed in\cite{cce142f4ee7344d48772f4406d19391d}.
Leveraging perceptual invariances in ``textures'', such as background hiss and noise, 
will likely result in simpler encodings.  Audio losses that are 
more perceptually-informed will undoubtedly yield more parsimonious encodings, 
as more model capacity will be spent on perceptually-relevant details.

\subsection{Encoder Variants}
\label{ssec:encoder_variants}

The anti-causal dilated convolutional network used in this iteration, 
while simple and parameter-efficient, may be suboptimal as an encoder.  
UNet and transformer enocder architectures are both worth exploring in 
future versions of this research.

\subsection{Decoder Variants}
\label{ssec:decoder_variants}

Anecdotally, we have observed that the roughly-physics-based 
inductive biases of the decoder encourage a sparser representation and 
better reconstruction quality, but many other decoder architectures are 
possible, with much to be learned from the larger field 
of differentiable digital signal processing\cite{engel2020ddspdifferentiabledigitalsignal}.

Another possibility is that events could be routed to specialized decoders, akin to the 
switch transformer architecture\cite{fedus2022switchtransformersscalingtrillion}.

\subsection{Redundant Events and Sparsity}
\label{ssec:decoder_variants}

Finally, aside from the physics-based inductive bias in the decoder, this work
makes no effort to further impose sparsity or penalize overly ``verbose'' representations.
It is our observation that this iteration of the model frequently produces redunant, 
duplicative events that could be collapsed further.

Future work should explore sparsity and/or energy penalties, seeking to maintain high 
reproduction quality while producing the sparsest, or lowest-energy representation
that can adequately explain the input signal.

\clearpage

\bibliographystyle{IEEEtran}
\bibliography{refs25}

\end{document}